\documentclass[10pt,twocolumn,letterpaper]{article}

\usepackage[pagenumbers]{cvpr} 
\usepackage{multirow}
\usepackage{pifont}
\newcommand{\xmark}{\ding{55}}

\definecolor{cvprblue}{rgb}{0.21,0.49,0.74}
\usepackage[pagebackref,breaklinks,colorlinks,allcolors=cvprblue]{hyperref}

\def\paperID{XXXX} 
\def\confName{CVPR}
\def\confYear{2026}

\title{RigMo: Unifying Rig and Motion Learning for Generative Animation}

\author{
Hao Zhang$^{1,2}$
Jiahao Luo$^{1,3}$
Bohui Wan$^{2}$
Yizhou Zhao$^{1,4}$
Zongrui Li$^{5}$ \\
\vspace{0.5em}
Michael Vasilkovsky$^{1}$
Chaoyang Wang$^{1}$
Jian Wang$^{1}$
Narendra Ahuja$^{2}$
Bing Zhou$^{1}$\\
$^{1}$Snap Inc. $^{2}$University of Illinois Urbana-Champaign $^{3}$University of California, Santa Cruz\\
$^{4}$Carnegie Mellon University $^{5}$Nanyang Technological University
}

\begin{document}
\maketitle

\begin{abstract}
Despite significant progress in 4D generation, rig and motion—the core structural and dynamic components of animation—are typically modeled as separate problems. Existing pipelines rely on ground-truth skeletons and skinning weights for motion generation and treat auto-rigging as an independent process, undermining scalability and interpretability.
We present \textbf{RigMo}, a unified generative framework that jointly learns rig and motion directly from raw mesh sequences, without any human-provided rig annotations. RigMo encodes per-vertex deformations into two compact latent spaces: a rig latent that decodes into explicit Gaussian bones and skinning weights, and a motion latent that produces time-varying SE(3) transformations. Together, these outputs define an animatable mesh with explicit structure and coherent motion, enabling feed-forward rig and motion inference for deformable objects.
Beyond unified rig–motion discovery, we introduce a Motion-DiT model operating in RigMo’s latent space and demonstrate that these structure-aware latents can naturally support downstream motion generation tasks.
Experiments on DeformingThings4D, Objaverse-XL, and TrueBones demonstrate that RigMo learns smooth, interpretable, and physically plausible rigs, while achieving superior reconstruction and category-level generalization compared to existing auto-rigging and deformation baselines. RigMo establishes a new paradigm for unified, structure-aware, and scalable dynamic 3D modeling. Project page: \url{https://RigMo-Page.github.io}
\end{abstract}

\section{Introduction}

Animation fundamentally couples structure and motion. The structure, typically represented by a rig, defines how an object can deform, while motion describes how that structure evolves over time. Yet despite their inherent interdependence, most existing pipelines treat these components in isolation.

Auto-rigging systems~\cite{Liao2022SkeletonFree,Song2021PoseTransfer} rely on artist-designed skeletons and skinning weights, training models to imitate human heuristics rather than understanding why certain rigs produce plausible deformations. As a result, these methods depend heavily on human annotations, which limits scalability and consistency across datasets and object categories.
A second class of motion-generation approaches assumes that a ground-truth rig is already provided. Human motion models~\cite{li2023example, tripathi2024humos, loper2023smpl, hong2025salad, yan2018spatial,aberman2020skeleton, zhong2023attt2m,zou2024parco,pinyoanuntapong2025maskcontrol,raab2023modi, yuan2023physdiff,tevet2022human} and animal- or object-centric pipelines~\cite{zuffi20173d, raab2023single} such as AnyTop~\cite{gat2025anytop} operate entirely in pose space, predicting joint rotations or SE(3) transforms on top of a predefined kinematic structure. These methods cannot infer rig structures, cannot handle arbitrary geometries, and fail when the assumed skeleton is mismatched or unavailable.
A third line of work removes rigging entirely. Modern vertex-space motion generators and 4D reconstruction methods~\cite{fu2024sync4d,wu2025animateanymesh,zhang2025gvf,zhang2024motiondiffuse,shi2025drive,bahmani20244d,bahmani2024tc4d,jiang2024animate3d}, including AnimateAnyMesh~\cite{wu2025animateanymesh} and GVFDiffusion~\cite{zhang2025gvf}, predict deformations frame by frame without any structural abstraction. While flexible, these models are difficult to control, hard to interpret, and unable to produce reusable animatable assets—the core purpose of rigging.
Together, these paradigms expose a fundamental gap: there is no unified framework that jointly learns rig structure and motion dynamics directly from raw mesh sequences, without predefined skeletons or per-sequence optimization. At the same time, relying on human-annotated rigs is costly and inconsistent, making a fully data-driven alternative both necessary and practical. This motivates a model capable of discovering structure from deformation and producing motion that conforms to that learned structure.

We introduce RigMo, a generative model that addresses this gap by jointly learning rig and motion with a unified architecture. RigMo encodes dynamic mesh sequences into two complementary latent spaces capturing spatial articulation and temporal evolution. The decoder reconstructs an explicit set of Gaussian bones~\cite{yang2021lasrlearningarticulatedshape}, from which skinning weights are derived, and predicts per-frame SE(3) transformations, together producing a fully animatable representation. This Gaussian-bone formulation provides a compact and scalable abstraction whose complexity depends only on the number of bones rather than mesh resolution and inherently resolution-insensitive. Furthermore, the model is trained in a self-supervised manner and requires no human annotations, enabling learning from unlabeled datasets such as DeformingThings4D~\cite{li20214dcomplete}.

In contrast to classical Smooth Skinning Decomposition with Rigid Bones (SSDR), which solves a separate optimization problem for each individual sequence, RigMo performs feed-forward rig and motion inference and generalizes across unseen categories and motion styles. The resulting representations are explicit, interpretable, and directly manipulable, supporting downstream applications such as motion generation, motion interpolation, and controllable editing. This naturally connects to our second contribution.

Beyond unified rig–motion discovery, we introduce Motion-DiT, a diffusion transformer operating in RigMo's motion latent space. By generating temporal trajectories in a structure-aware latent domain rather than raw vertex coordinates, Motion-DiT provides a principled pathway toward controllable, high-quality motion synthesis and demonstrates the utility of RigMo’s learned latent representation for downstream motion-generation tasks.

Across DeformingThings4D, Objaverse-XL~\cite{deitke2023objaverse}, and TrueBones, RigMo recovers physically plausible skeletons, coherent skinning structures, and realistic motions directly from raw deformation data, establishing a unified, structure-aware, and controllable framework for 4D generative animation.

\begin{figure*}[t]
    \centering
    \includegraphics[width=0.9\linewidth]{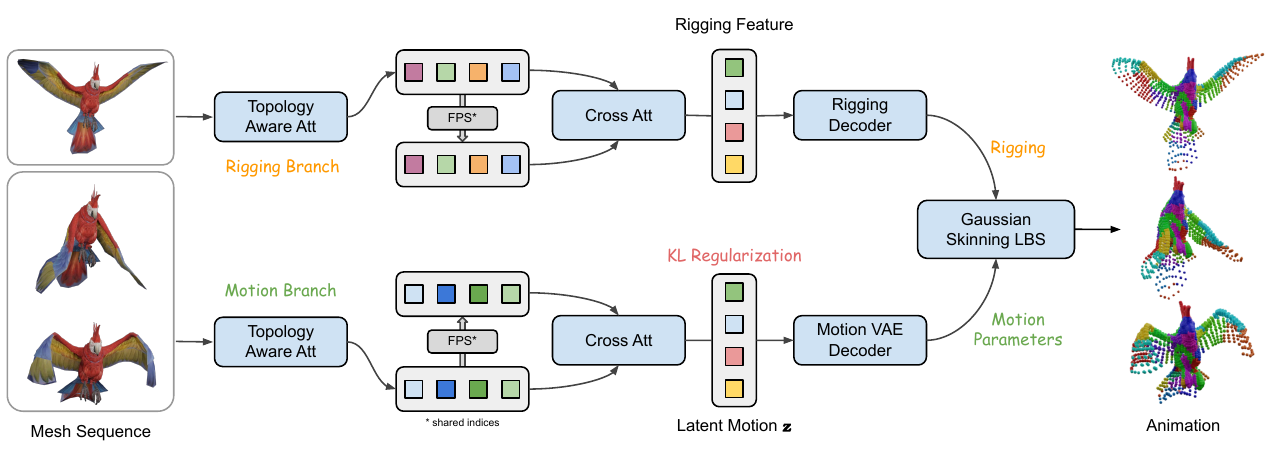}
    \vspace{-5pt}
    \caption{
        Overview of the \textbf{RigMo-VAE} framework.
        Given temporal vertex trajectories from deforming mesh sequences, RigMo employs a dual-path encoder to disentangle static geometry (rigging branch) and dynamic motion (motion branch), learning a compact latent representation that captures both spatial structure and temporal dynamics.
        The decoder maps these latent features to physically interpretable rig components: Gaussian bone descriptors defining geodesic-aware skinning weights and variational motion parameters for local and root transformations.
        Different colors indicate the influence regions of learned Gaussian bones, demonstrating semantically meaningful decomposition of mesh deformation without manual rigging supervision.
    }
    \label{fig:overview}
    \vspace{-5pt}
\end{figure*}
\section{Related Works}

\textbf{4D Generative Models.} Recent work has produced a variety of generative models for 4D generation~\cite{jiang2023consistent4d, zeng2024stag4d, cao2024motion2vecsets, wu2024sc4d, zhang2024magicpose4d, zhang20244diffusion, li2024dreammesh4d, fu2024sync4d, zhu2025ar4d, yao2025sv4d, xie2024sv4d, chen2025v2m4, ren2024l4gm, wang2024animatabledreamer}. A common framework is the use of carefully designed 3D or 4D variational autoencoders to embed geometry and motion into latent spaces, with generative networks (\ie, diffusion models~\cite{ho2020denoising}) decoding these latents conditioned on texts or videos. Although such methods yield visually compelling results, the resulting motions often lack controllability: they produce animations without explicit rigging and corresponding motion parameters, which greatly limits their applicability in practical use. In this work, we address this issue by encoding deformable meshes into unified spatio-temporal latent representations that can be further decoded into highly controllable rig structures and associated motions.

\medskip
\noindent\textbf{Auto-Rigging.}
Auto-rigging aims to convert a static mesh into an animatable asset by predicting a skeleton and per-vertex skinning weights. 
Template-based approaches~\cite{Baran2007Pinocchio,ma2023tarig,li2021learning} attach predefined skeletal structures and optimize their poses and weights, but generalize poorly beyond humanoid shapes. 
Template-free methods such as RigNet~\cite{rignet} relax assumptions on joint topology, though they may produce overly dense skeletons or rely on handcrafted geometric cues. 
More recently, auto-regressive models~\cite{liu2025riganything,song2025puppeteer,zhang2025unirig,song2025magicarticulate} treat skeleton generation as a sequence prediction problem, leveraging large datasets such as Articulation-XL~\cite{song2025magicarticulate} to learn category-agnostic articulation priors. 
Despite progress, these approaches still depend on artist-annotated rigs at scale, which remain costly and difficult to obtain.

\medskip
\noindent\textbf{Optimization-Based Inverse Skinning.}
Inverse skinning~\cite{zhanglearning,sun2024ponymation,wu2023magicponylearningarticulated3d,zhang2024s3o,zhang2025physrig} considers the complementary task to auto-rigging: given a sequence of articulated mesh deformations, the objective is to recover the underlying rig parameters, including bone transformations, skinning weights, or elasticity properties.  
Classical solutions based on Linear Blend Skinning (LBS) offer computational efficiency but restrict deformation to a linear blend of transformations, limiting realism.  
Recent physically grounded formulations~\cite{zhang2025physrig} jointly optimize skeletal motion and material properties, enabling non-linear volumetric behavior.
However, these pipelines require \emph{per-case} nonlinear optimization, making them slow and difficult to scale.  
Moreover, the recovered rigs are tightly bound to individual sequences and fail to generalize across subjects or motion styles.  
In contrast, our method predicts both motion trajectories and implicit rig parameters in a single feed-forward pass, jointly learning from all training sequences to obtain a shared, generalizable articulation representation that transfers robustly across diverse objects and motions.

\section{Method}

\textbf{RigMo} is a novel framework for modeling deformable surface motion by learning rigging representations composed of Gaussian bones (see Gaussian Bone Definition in Sec.~\ref{sec:gaussian_bone_def}) $\mathbf{G}$ and their associated motion parameters $\{\mathbf{q}_{\text{local}}, \mathbf{t}_{\text{local}}, \mathbf{q}_{\text{root}}, \mathbf{t}_{\text{root}}\}$. Here $\mathbf{q}$ denotes rotation quaternions and $\mathbf{t}$ represents translation vectors, which jointly govern mesh deformation through differentiable skinning. Unlike SSDR methods that require per-case fitting, RigMo is a feedforward framework capable of inferring rig structure and motion directly from input sequences.

Our method consists of two main components: the \textbf{RigMo VAE} (Sec.~\ref{sec:rigmo_vae}), which learns the rigging representation and corresponding motion parameters, and the \textbf{Motion DiT} module (Sec.~\ref{sec:motion_dit}), which operates in RigMo VAE’s learned latent space and uses rig-branch outputs as conditioning signals to diffuse motion latents for controllable animation generation.

\subsection{RigMo VAE}
\label{sec:rigmo_vae}

Given a mesh sequence $\mathbf{V}\in\mathbb{R}^{B\times T\times N\times 3}$, with batch size $B$, frame count $T$, and $N$ vertices per mesh, RigMo reconstructs the deformable motion $\hat{\mathbf{V}}\in\mathbb{R}^{B\times (T-1)\times N\times 3}$ from latent representations that encode both spatial structure and temporal dynamics.
The network architecture consists of:
(1) a topology-aware encoder that aggregates geometric and motion cues,
(2) a decoder that predicts Gaussian bones and their transformations, and
(3) a Gaussian-based Linear Blend Skinning (LBS) module that synthesizes deformations under learned rig parameters.

\subsubsection{Topology-Aware Encoder}

RigMo's encoder jointly captures static geometry and dynamic motion through a dual-path architecture designed to disentangle canonical shape information from temporal deformation cues. This ensures that rigging prediction reflects stable structure rather than specific motion instances, enabling generalizable, consistent, and interpretable bone–vertex relationships.

\paragraph{Rigging Branch.}
The rigging branch processes the canonical mesh geometry (first frame) to establish bone–vertex correspondence and predict Gaussian bone parameters. We encode the input vertices 
$\mathbf{V}_0 \in \mathbb{R}^{N \times 3}$ 
into per-vertex embeddings $\mathbf{V}_0^{\text{emb}}$ via topology-aware attention:
\begin{equation}
    \mathbf{h}_{\ell} = \text{Attn}(\text{LN}(\mathbf{h}_{\ell-1}), \mathcal{N}) + \mathbf{h}_{\ell-1}.
\end{equation}
Farthest Point Sampling selects $K$ bone tokens 
$\mathbf{B}^{\text{emb}}$ and coordinates $\mathbf{C}_{\text{bone}}$.
Cross-attention produces bone–vertex correlation features:
\begin{equation}
    \mathbf{A}_{\text{rig}} = \text{CrossAttn}(\mathbf{B}^{\text{emb}}, \mathbf{V}_0^{\text{emb}}, \mathbf{V}_0^{\text{emb}}),
\end{equation}
from which each bone predicts Gaussian parameters 
$\mathbf{G}_k = [\Delta \mathbf{c}_k, \mathbf{s}_k, \mathbf{q}_k]$.

\paragraph{Motion Branch.}
The motion branch encodes temporal deformation patterns and predicts variational latent variables governing local and global motion.
We first compute per-frame vertex displacements:
\begin{equation}
    \mathbf{V}_{\Delta} = \mathbf{V}[:,1:,:,:]-\mathbf{V}[:,:-1,:,:]
    \in \mathbb{R}^{B \times (T-1) \times N \times 3}.
\end{equation}
These displacements are processed through temporal–spatial attention layers to obtain
$\mathbf{V}_{\Delta}^{\text{emb}}$.
Using the shared bone token coordinates $\mathbf{C}_{\text{bone}}$, we extract bone–motion interaction features:
\begin{equation}
    \mathbf{A}_{\text{motion}}
    = \text{CrossAttn}(\mathbf{B}^{\text{emb}}, \mathbf{V}_{\Delta}^{\text{emb}}, \mathbf{V}_{\Delta}^{\text{emb}})
    \in \mathbb{R}^{B \times (T-1) \times K \times d_b}.
\end{equation}

These features are fed into the local-motion posterior estimator, which predicts bone-level posterior parameters for each bone $k$ and frame $t$:
\begin{equation}
    [\boldsymbol{\mu}_{\text{local}}, \log \boldsymbol{\sigma}_{\text{local}}]
    = \text{MLP}(\mathbf{A}_{\text{motion}}).
\end{equation}
Latent variables are then sampled via:
\begin{equation}
    \mathbf{z}_{\text{local}}
    = \boldsymbol{\mu}_{\text{local}}
    + \boldsymbol{\sigma}_{\text{local}} \odot \boldsymbol{\epsilon},
    \quad \boldsymbol{\epsilon}\sim\mathcal{N}(0,\mathbf{I}).
\end{equation}

For the global transformation, a global-motion posterior estimator temporally aggregates motion features 
(average pooling), producing root-level posterior parameters:
\begin{equation}
    [\boldsymbol{\mu}_{\text{root}}, \log \boldsymbol{\sigma}_{\text{root}}]
    = \text{MLP}(\text{Agg}(\mathbf{A}_{\text{motion}})).
\end{equation}
A root-level latent code is sampled as:
\begin{equation}
    \mathbf{z}_{\text{root}}
    = \boldsymbol{\mu}_{\text{root}}
    + \boldsymbol{\sigma}_{\text{root}} \odot \boldsymbol{\epsilon}.
\end{equation}

\subsubsection{Rig-Motion Decoder}

The decoder maps the latent codes to physically meaningful transformations, comprising a static Gaussian-bone decoder and motion decoders for local and root $\mathrm{SE}(3)$ transformations. Taking as input the latent features produced by the rigging branch, a lightweight MLP-based module predicts soft Gaussian bone regions
$\mathbf{G}=[\Delta\mathbf{c},\mathbf{s},\mathbf{q}]$ for each bone, while local and global motion are decoded from the latent variables predicted in the motion encoder. Concretely, a local-motion decoder maps $\mathbf{z}_{\text{local}}$ to per-bone transformations:
\begin{equation}
    \{\mathbf{q}_{\text{local}}, \mathbf{t}_{\text{local}}\}
    = \text{Dec}_{\text{local}}(\mathbf{z}_{\text{local}}).
\end{equation}
A separate root-motion decoder maps $\mathbf{z}_{\text{root}}$ to global transformations:
\begin{equation}
    \{\mathbf{q}_{\text{root}}, \mathbf{t}_{\text{root}}\}
    = \text{Dec}_{\text{root}}(\mathbf{z}_{\text{root}}).
\end{equation}





\begin{figure*}[t]
    \centering
    \includegraphics[width=0.75\linewidth]{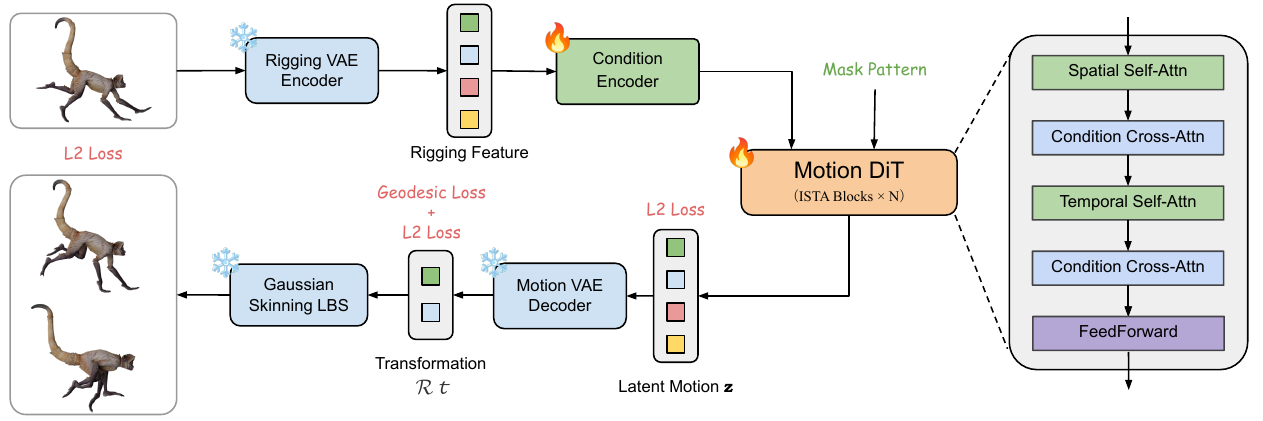}
    \vspace{-5pt}
    \caption{
        Overview of the \textbf{Motion DiT}. Given static rigging features, a condition encoder produces anchor and global tokens that guide a diffusion transformer operating in RigMo’s motion-latent space. The model uses spatial, temporal, and frame-conditioned cross-attention to predict denoised motion latents, which are decoded into bone transformations and vertex sequences via Gaussian skinning.
    }
    \vspace{-5pt}
    \label{fig:overview_dit}
\end{figure*}

\subsubsection{Gaussian Skinning LBS Module}

\paragraph{Gaussian Bone Definition.}
\label{sec:gaussian_bone_def}
Each Gaussian bone $k$ is defined by its geometric parameters $\mathbf{G}_k = [\mathbf{c}_k, \mathbf{s}_k, \mathbf{q}_k]$, where $\mathbf{c}_k \in \mathbb{R}^3$ represents the bone center, $\mathbf{s}_k \in \mathbb{R}^3$ denotes the anisotropic scaling factors, and $\mathbf{q}_k \in \mathbb{R}^4$ is the orientation quaternion.
These parameters collectively define a 3D Gaussian ellipsoid that acts as a soft bone with spatially-varying influence.

\paragraph{Skinning Weight Computation.}
For each vertex $\mathbf{v}_i$, the skinning weight with respect to bone $k$ is computed using the Mahalanobis distance within the Gaussian bone's coordinate system:
\begin{equation}
    w_{ik}^{\text{raw}} = 
    \frac{
        \exp\!\left(-\tfrac{1}{2}\|\mathbf{R}_k^\top (\mathbf{v}_i - \mathbf{c}_k) \oslash \mathbf{s}_k\|^2\right)
    }{
        \sum_{j=1}^K \exp\!\left(-\tfrac{1}{2}\|\mathbf{R}_j^\top (\mathbf{v}_i - \mathbf{c}_j) \oslash \mathbf{s}_j\|^2\right)
    },
\end{equation}
where $\mathbf{R}_k$ is the rotation matrix derived from quaternion $\mathbf{q}_k$, and $\oslash$ denotes element-wise division.

\paragraph{Linear Blend Skinning.}
Given the skinning weights and bone transformations, each vertex is deformed through Linear Blend Skinning:
\begin{equation}
    \hat{\mathbf{v}}_i = \sum_{k=1}^K w_{ik} \cdot \mathbf{T}_k \cdot \tilde{\mathbf{v}}_i,
\end{equation}
where $\mathbf{T}_k = \mathbf{T}_{\text{root}} \cdot \mathbf{T}_{k,\text{local}}$ represents the hierarchical transformation combining root motion and local bone motion, and $\tilde{\mathbf{v}}_i$ is the vertex in homogeneous coordinates.

\paragraph{Topological Coherence via Geodesic-Aware Weight Refinement.}
While Gaussian weighting captures local proximity, it may erroneously couple topologically distant regions that are spatially close (e.g., arm–torso contact).
To mitigate such artifacts, we introduce a \emph{geodesic-aware weight refinement} strategy that enforces topological consistency on the mesh surface.

Let $V=\{v_1,\dots,v_N\}$ denote vertices and $A=\{a_1,\dots,a_K\}$ denote bone anchors with raw weights $W^{\mathrm{raw}}\in\mathbb{R}^{N\times K}$.
We define the surface geodesic distance between vertex $v_i$ and anchor $a_k$ as:
\begin{equation}
    d_g(v_i,a_k)=\min_{\pi\in \Pi(a_k,v_i)}\sum_{(v_p,v_q)\in \pi}\lVert v_p-v_q\rVert_2,
\end{equation}
where $\Pi(a_k,v_i)$ enumerates edge-connected vertex paths.
A binary coherence mask is constructed as
\begin{equation}
    M_{ik}=
    \begin{cases}
        1, & d_g(v_i,a_k)<\tau,\\
        0, & \text{otherwise.}
    \end{cases}
\end{equation}
The refined skinning weights become:
\begin{equation}
    \tilde{W}_{ik}=W^{\mathrm{raw}}_{ik}M_{ik}, \quad
    w_{ik}=\frac{\tilde{W}_{ik}}{\sum_{j=1}^K \tilde{W}_{ij}+\varepsilon},
\end{equation}
with $\varepsilon=10^{-8}$ for numerical stability.
Vertices with no reachable bones are assigned a one-hot weight to their nearest bone.
This refinement effectively suppresses cross-part influence and yields cleaner, more coherent skinning, with vertices typically retaining influence from only 2–3 bones.

\paragraph{Self-Supervised Training Objectives.}
A key advantage of RigMo is that it requires no rigging annotations or supervision during training.
Instead, the model learns underlying rig structures in a purely \emph{self-supervised} manner directly from motion data alone.
This eliminates the need for expensive manual rigging annotations from skilled artists, making the approach highly scalable to large-scale motion datasets.
RigMo-VAE is trained end-to-end using only two self-supervised objectives:
a vertex-level reconstruction loss and a latent regularization term:
\begin{equation}
    \mathcal{L}_{\text{total}}
    = \lambda_{\text{recon}}\mathcal{L}_{\text{recon}}
    + \lambda_{\text{KL}}\mathcal{L}_{\text{KL}}.
\end{equation}
The reconstruction term supervises per-vertex motion fidelity without requiring any rig supervision:
\begin{equation}
    \mathcal{L}_{\text{recon}}
    = \frac{1}{BTN}\sum_{b,t,i}\|\hat{\mathbf{v}}_{b,t,i}-\mathbf{v}_{b,t,i}\|^2,
\end{equation}
while the KL divergence regularizes latent distributions toward a unit Gaussian prior:
\begin{equation}
    \mathcal{L}_{\text{KL}}
    = \frac{1}{2}\sum_i(\mu_i^2+\sigma_i^2-\log\sigma_i^2-1).
\end{equation}
This pure motion-driven formulation enables the model to discover semantically meaningful bone structures and their relationships purely from observing vertex trajectories, bypassing the traditional bottleneck of manual rigging annotation and enabling scalable learning from diverse motion collections.

\begin{figure*}[t]
    \centering
    \includegraphics[width=\linewidth]{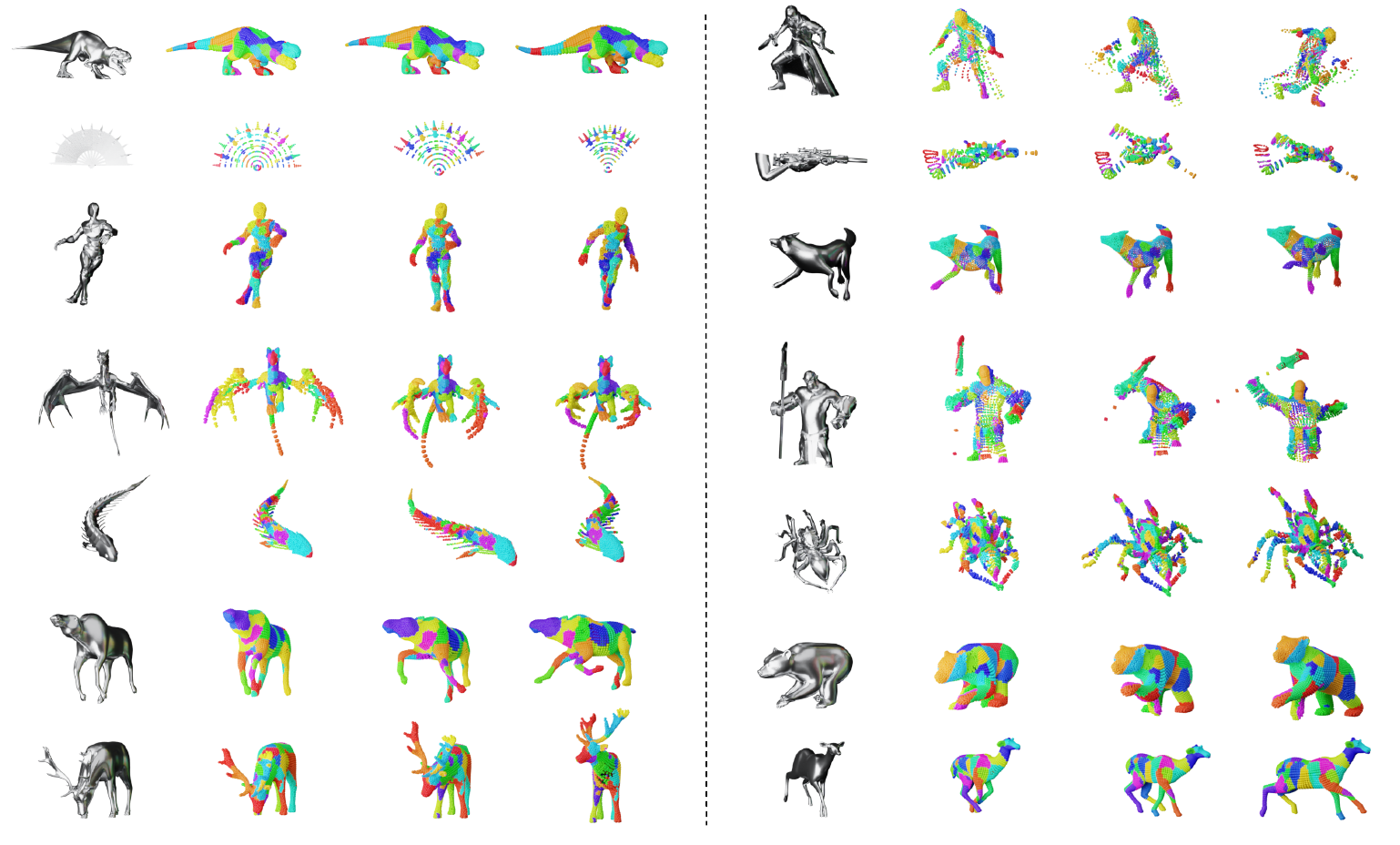}
    \vspace{-20pt}
    \caption{Results produced by the full RigMo. Given a sparse input sequence, where a subset of frames is observed according to a frame mask, RigMo reconstructs a complete animatable model by jointly predicting the rigging structure (Gaussian bones and skinning weights) and synthesizing the missing motion frames through diffusion in the RigMo latent space. The resulting rigged model produces coherent, articulated motion across humans, animals, and diverse non-human shapes, demonstrating that sparse observations are sufficient to recover a full animation without category-specific priors.}
    \label{fig:3}
\end{figure*}

\subsection{Motion DiT}
\label{sec:motion_dit}

Given static rigging features as conditions, the Motion DiT generates motion latents in the RigMo-VAE space.
A condition encoder aggregates static/rigging cues (e.g., anchor/gaussian/skinning features) into
anchor tokens $\mathbf{A}\!\in\!\mathbb{R}^{B\times K\times H}$ and a global token $\mathbf{g}\!\in\!\mathbb{R}^{B\times H}$,
which remain fixed during generation and guide the denoising process.
We first project the VAE’s dynamic tokens $\mathbf{Z}_{\text{dyn}}\!\in\!\mathbb{R}^{B\times K\times(T{-}1)\times D_{\text{dyn}}}$ and
root tokens $\mathbf{Z}_{\text{root}}\!\in\!\mathbb{R}^{B\times(T{-}1)\times D_{\text{root}}}$ to a common width $H$
and form a unified motion-latent tensor
$\mathbf{Z}_{\text{mot}}\!\in\!\mathbb{R}^{B\times (K{+}1)\times (T{-}1)\times H}$ by concatenating the $K$ bone streams with the root stream.
Conditioned on a configurable frame-mask schedule that specifies observed and generated frames, the diffusion transformer predicts the velocity fields $\hat{\mathbf{v}}_{\text{mot}}$ for the generated frames, while keeping the observed frames fixed as conditioning signals.
$\hat{\mathbf{x}}_0$ is then recovered via $v$-prediction and decoded by RigMo-VAE decoder to animations.
This conditioning-by-statics and generation-in-motion-latents design ties the synthesized dynamics tightly to the learned rig structure.

The backbone consists of $L{=}12$ interleaved spatio-temporal attention (ISTA) blocks with hidden dimension $H{=}512$.  
Each block alternates spatial attention (within frame, across bones) and temporal attention (within bone, across frames), augmented by two conditioning pathways:  
(1) cross-attention injecting static and global priors ($\mathbf{A},\mathbf{g}$), and  
(2) frame-level cross-attention that aligns generated frames with observed ones based on masks.  
The network outputs the velocity parameterization $\hat{\mathbf{v}}$, from which the denoised latent state is reconstructed as  
$\hat{\mathbf{x}}_0=\sqrt{\alpha_t}\mathbf{x}_t-\sqrt{1-\alpha_t}\hat{\mathbf{v}}$.  
Losses are computed only on generated frames:
\[
\mathcal{L}
= \lambda_{\text{lat}} \mathcal{L}_{\text{lat}}
+ \lambda_{\text{rot}} \mathcal{L}_{\text{rot}}
+ \lambda_{\text{trans}} \mathcal{L}_{\text{trans}}
+ \lambda_{\text{vert}} \mathcal{L}_{\text{vert}},
\]
where $\mathcal{L}_{\text{lat}}$ denotes the latent-space L2 loss, $\mathcal{L}_{\text{rot}}$ the SO(3) geodesic rotation loss, $\mathcal{L}_{\text{trans}}$ the translation L2 loss, and $\mathcal{L}_{\text{vert}}$ the vertex-space L2 loss.
Weights are set to $(\lambda_{\text{lat}},\lambda_{\text{rot}},\lambda_{\text{trans}},\lambda_{\text{vert}})=(0.5,1.0,0.2,0.1)$.  
This design enables Motion DiT to generate or interpolate complex, long-range motions in RigMo’s latent space while maintaining spatial coherence and temporal realism.

\section{Experiments}
Our primary focus in this paper is the proposed RigMo-VAE, which forms the core of our rig–motion representation. 
While we additionally introduce a Motion DiT to demonstrate the effectiveness of RigMo-VAE as a latent space for downstream motion generation, the main contributions of this work lie in the rigging and motion decomposition capabilities of the VAE itself. 
Due to space constraints, more results for the Motion DiT are provided in the supplementary material.

\begin{figure*}[t]
    \centering
    \includegraphics[width=\linewidth]{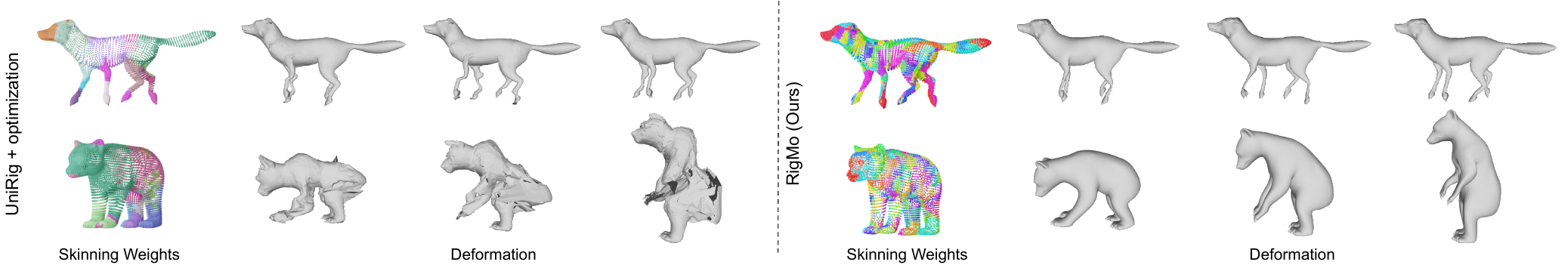}
    \vspace{-15pt}
    \caption{
    Comparison between UniRig+Optimization and our RigMo Rigging Module.
    Although UniRig may produce visually plausible skinning weights in some cases (e.g., the fox),
    its rigging does not generalize and collapses under actual animation, leading to severe deformation artifacts.
    In contrast, RigMo learns robust and transferable rig structures \emph{directly from motion}, without any ground-truth rig supervision, and achieves stable, high-fidelity deformations across diverse poses and animal species.
    }
    \vspace{-5pt}
    \label{fig:4}
\end{figure*}

\subsection{Datasets}
We curate a large-scale corpus of $\sim$20{,}000 deformable mesh sequences spanning three complementary domains to ensure diverse motion coverage and strong generalization.  
\textbf{DeformingThings4D~\cite{li20214dcomplete}} provides 1{,}972 real-world sequences capturing organic non-rigid deformations, where sequences shorter than 10 frames are excluded for temporal stability.  
\textbf{TrueBones~\cite{lee2025move}} contributes 1{,}287 high-fidelity articulated animations with realistic skeletal motion.  
\textbf{Objaverse-XL~\cite{deitke2023objaverse}} offers 17{,}024 synthetic sequences (24 frames each) filtered through a motion-quality classifier to retain diverse topologies and motion styles across categories.
For each dataset, we adopt a 5:1 split into training and test sets to enable consistent evaluation across domains.

To handle heterogeneous mesh resolutions while preserving geometric consistency, we adopt a topology-preserving vertex normalization protocol.  
Meshes exceeding 20K vertices are downsampled to 5K via Farthest Point Sampling (FPS) with geodesic neighborhood preservation, while lower-resolution meshes are iteratively subdivided until surpassing the target vertex count, then downsampled identically.  
This procedure yields uniform vertex embeddings with consistent local connectivity, facilitating robust and topology-aware rigging discovery.

\begin{table*}[t]
\centering
\scalebox{0.82}{%
\begin{tabular}{l|cc|cc|cc}
\toprule
\multirow{2}{*}{Method} & \multicolumn{2}{c|}{Training Reconstruction} & \multicolumn{2}{c|}{Cross-Motion Transfer} & \multicolumn{2}{c}{Mean $\downarrow$} \\
& CD-L1 $\downarrow$ & CD-L2 $\downarrow$ & CD-L1 $\downarrow$ & CD-L2 $\downarrow$ & CD-L1 $\downarrow$ & CD-L2 $\downarrow$ \\
\midrule
Per-Case Optimization & 12.3 $\pm$ 0.2 & 8.21 $\pm$ 0.3 & 68.8 $\pm$ 6.7 & 43.5 $\pm$ 5.4 & 40.55 & 25.86 \\
UniRig~\cite{zhang2025unirig} + Optimization & 37.3 $\pm$ 2.3 & 28.4 $\pm$ 2.1 & 48.6 $\pm$ 4.9 & 31.2 $\pm$ 4.6 & 42.95 & 29.80 \\
MagicArticulate~\cite{song2025magicarticulate} + Opt. & 43.1 $\pm$ 3.7 & 23.9 $\pm$ 3.2 & 53.4 $\pm$ 5.2 & 28.7 $\pm$ 3.9 & 48.25 & 26.30 \\
\midrule
\textbf{RigMo (Ours)} & \textbf{11.1} $\pm$ \textbf{0.6} & \textbf{7.64} $\pm$ \textbf{0.3} & \textbf{13.82} $\pm$ \textbf{0.49} & \textbf{11.83} $\pm$ \textbf{0.34} & \textbf{12.46} & \textbf{9.74} \\
\bottomrule
\end{tabular}%
}
\vspace{-3pt}
\caption{Rigging Discovery and Cross-Motion Generalization ($\times 10^{-3}$). Lower is better. Experiments are done on DT4D test dataset.}
\label{tab:inverse_skinning}
\vspace{-5pt}
\end{table*}

\subsection{Implementation Details}

\textbf{RigMo-VAE} is implemented in PyTorch and trained with mixed precision on 24$\times$A100 (80\,GB) GPUs. 
Each batch contains 144 mesh sequences ($T{=}20$, $N{=}5$K). 
We use AdamW ($\beta_1{=}0.9$, $\beta_2{=}0.999$, weight decay $10^{-4}$) with a learning rate of $3{\times}10^{-4}$, 2K warm-up steps, cosine decay, gradient clipping (1.0), and EMA (0.999). 
The encoder uses six topology-aware attention layers (hidden dim 256, eight heads, neighborhood size $k{=}5$). 
Anchor sampling selects $K{=}48/128$ Gaussian bones, and skinning weights are sparsified to the top $K_s{=}4$. 
Training uses reconstruction and KL losses ($\lambda_{\text{recon}}{=}1.0$, $\lambda_{\text{KL}}{=}10^{-6}$), with KL annealed over the first 30\% of training.
On the full corpus, RigMo converges after $\sim$50K steps ($\sim$10 days). 
Training only on DeformingThings4D reaches comparable reconstruction within $\sim$7.5K epochs ($\sim$1.5 days). 
Inference is fully feed-forward, reconstructing a 20-frame 5K-vertex sequence in $\sim$40\,ms per frame on an A100 GPU.

\noindent\textbf{Motion DiT} is implemented in PyTorch with sinusoidal timestep embeddings and a 12-block ISTA backbone applied to 512-d motion tokens. 
We adopt a DDPM scheduler with 1000 steps and scaled-linear $\beta$, using the $v$-prediction parameterization. 
Training uses AdamW ($\beta_1{=}0.9$, $\beta_2{=}0.999$, weight decay $0.01$) with an initial learning rate of $1{\times}10^{-4}$ and cosine annealing on 4 H200 GPUs (batch size 64). 
End-to-end training takes $\sim$40 hours and converges within $\sim$5000 epochs. 
The model remains stable without auxiliary losses, aided by its spatial–temporal attention design and sparse conditioning in RigMo's motion-latent space.

\subsection{Rigging Evaluation}

\textbf{Evaluation Protocol.}
We conduct comprehensive rigging evaluation on the DeformingThings4D partition, leveraging its annotation-free nature to assess genuine unsupervised rig-discovery capabilities. Our experimental protocol randomly samples pairs of motion sequences from the same object’s training data and testing data, forming the training and testing splits used for evaluation, with training sequences used for baseline optimization and testing sequences used to assess cross-motion rig transferability. Statistical significance is ensured through 100 independent random splits, reporting averaged metrics and confidence intervals.

\noindent\textbf{Competitive Baselines.}
As shown in Tab.~\ref{tab:inverse_skinning}, we establish rigorous comparisons against two paradigmatic approaches representing current practice:
\textbf{Per-Case Optimization.}  
This baseline jointly optimizes rig parameters $\{\mathbf{G}, \mathbf{W}\}$ and motion parameters $\{\mathbf{T}_t\}$ via gradient descent on the reconstruction loss $\mathcal{L}_{\text{recon}}$.  
During testing, rig parameters remain fixed while only transformations are optimized, directly measuring generalization capacity.  
However, because the predicted rigging is fully determined by the training sequence itself, this approach can only perform well on the specific cases it was optimized for. When applied to unseen motions, the optimized rig often collapses or becomes invalid, revealing a lack of generalization.
\textbf{Auto-Rigging Pipeline.}  
We employ state-of-the-art automatic rigging methods (UniRig~\cite{zhang2025unirig}, MagicArticulate~\cite{song2025magicarticulate}) to generate initial bone structures from canonical poses, followed by per-sequence transformation optimization. This pipeline reflects the current industry standard for automated character rigging.

As illustrated in Fig.~\ref{fig:4}, the generalization ability of auto-rigging pipelines is also constrained by the diversity and structure of the training data. In many cases, these methods produce bone layouts that appear visually plausible, yet they fail to reproduce even common motions during animation. Importantly, even experienced artists cannot rely purely on visual priors to assign correct skinning weights; iterative adjustment is always required to produce a rig that accurately reproduces target deformations. This observation is consistent with the core motivation behind RigMo—robust rigging cannot be inferred from static geometry alone and must be learned directly from motion.

\begin{table}[t]
\centering
\scalebox{0.8}{%
\begin{tabular}{l|cc|c}
\toprule
\multirow{2}{*}{Method} & \multicolumn{2}{c|}{Geometric Accuracy} & \multirow{2}{*}{Time (20f) $\downarrow$} \\
& CD-L1 $\downarrow$ & CD-L2 $\downarrow$ & \\
\midrule
AnimateAnyMesh~\cite{wu2025animateanymesh} & 1.81 $\pm$ 0.13 & 1.32 $\pm$ 0.01 & 2.8s \\
Step1X3D~\cite{li2025step1x} & 3.63 $\pm$ 0.21 & 2.96 $\pm$ 0.18 & 22.6s \\
Hunyuan3D~\cite{zhao2025hunyuan3d} & 3.21 $\pm$ 0.19 & 2.67 $\pm$ 0.14 & 17.4s \\
\midrule
\textbf{RigMo (Ours)} & \textbf{1.73} $\pm$ \textbf{0.11} & \textbf{1.26} $\pm$ \textbf{0.08} & \textbf{0.74s} \\
\bottomrule
\end{tabular}%
}
\vspace{-5pt}
\caption{Reconstruction fidelity and inference efficiency ($\times 10^{-2}$ for CD). 
RigMo achieves the best geometric accuracy and the fastest inference among all baselines.}
\vspace{-5pt}
\label{tab:vae_reconstruction}
\end{table}

\subsection{RigMo-VAE Reconstruction Evaluation}

We establish comprehensive reconstruction benchmarks against leading mesh-generation architectures on our curated sub–test set of 500 sequences spanning diverse deformation patterns and object categories.

We compare against representative paradigms spanning both sequence-aware and frame-independent generative models (as shown in Tab.~\ref{tab:vae_reconstruction}):
\textbf{AnimateAnyMesh}~\cite{wu2025animateanymesh} employs a sequence-aware VAE architecture for temporal mesh modeling.  
However, its design requires a high-dimensional 512-token representation to encode object motion, whereas RigMo-VAE achieves superior reconstruction quality using only 48/128 tokens, resulting in significantly faster inference and reduced memory footprint.
\textbf{Step1X3D}~\cite{li2025step1x} and \textbf{Hunyuan~3D~2.1}~\cite{zhao2025hunyuan3d} represent state-of-the-art frame-independent 3D VAEs that perform per-frame generation, incurring substantial computational overhead for sequence reconstruction.  
These methods inherently break vertex correspondence across frames and tend to lose fine-grained surface details due to their framewise decoding process.  
Because their latent spaces operate on inconsistent canonicalizations, our evaluation applies a pose-invariant correction: we record the canonical transformations $\{\mathbf{R}, \mathbf{t}, s\}$ used for input normalization and apply the inverse transformations after reconstruction to enable fair, shape-focused geometric assessment.

\subsection{Qualitative Analysis}

As shown in Fig.\ref{fig:3}, we present comprehensive visualizations demonstrating RigMo's semantic understanding and generation capabilities across three critical aspects:
\textbf{Learned Rigging Structures.} Our visualizations reveal semantically meaningful bone-vertex correspondences emerging without supervision, with distinct Gaussian bones capturing intuitive anatomical regions (limbs, torso, extremities) across diverse object categories.
\textbf{Controllable Motion Synthesis.} Motion DiT demonstrations exhibit precise conditional generation capabilities, producing plausible animations that respect learned rig constraints while enabling creative exploration of novel motion patterns.

\subsection{Diagnostics}


\begin{table}[t]
\centering
\scalebox{0.9}{%
\begin{tabular}{l|cc}
\toprule
Configuration & CD-L1 $\downarrow$ & CD-L2 $\downarrow$ \\
\midrule
w/o Geodesic Refinement & 2.37 $\pm$ 0.15 & 2.07 $\pm$ 0.18 \\
48 Bone Tokens & 1.91 $\pm$ 0.13 & 1.48 $\pm$ 0.12 \\
128 Bone Tokens & \textbf{1.73} $\pm$ \textbf{0.11} & \textbf{1.26} $\pm$ \textbf{0.08}  \\
\bottomrule
\end{tabular}%
}
\vspace{-5pt}
\caption{Ablation study on the DeformingThings4D validation set using CD-L1/L2 metrics ($\times 10^{-2}$).}
\label{tab:ablation}
\vspace{-5pt}
\end{table}

\noindent\textbf{Resolution-Agnostic Architecture.}
Regardless of the original resolution, all meshes are first resampled to a 5K-vertex representation for processing in the encoder.  
Since Gaussian bones and motion transformations are defined continuously in 3D space, rather than tied to specific vertex indices, the predicted rig and motion parameters can be directly applied back to the original mesh resolution.
This property results in a consistent deformation quality even when the mesh tessellation changes, as the learned bones capture geometric regions rather than fixed vertex subsets.    
This contrasts with mesh-dependent approaches~\cite{wu2025animateanymesh}, where changing vertex count typically alters the learned skinning structure and degrades animation fidelity.

\noindent\textbf{Topological Refinement Impact.}
Next, we evaluate the effectiveness of our geodesic-aware weight refinement module.  
Removing this component leads to a substantial decline in reconstruction quality as shown in Tab.\ref{tab:ablation}.   
These results highlight that spatial proximity alone is insufficient for articulated objects, where vertices may be close in Euclidean space but distant in the underlying kinematic topology.

\noindent\textbf{Bone Token Cardinality Trade-off.}
Finally, we analyze the effect of bone token cardinality.  
While increasing from 48 to 128 tokens yields a 0.018\% improvement in CD-L1, the gains diminish as tokens become overly fine-grained.  
Although 128 tokens achieve the best quantitative reconstruction, 48 tokens provide a more favorable balance between efficiency, interpretability, and stability.  
Excessive tokens tend to fragment coherent anatomical regions without offering proportional improvements.

\section{Conclusion}

In this paper, we introduce RigMo, a unified generative framework that jointly learns rig structure and motion dynamics directly from raw mesh sequences without any rig annotations. RigMo factorizes per-vertex deformations into two compact latent spaces that decode into explicit Gaussian bones, skinning weights, and bone transformations, producing fully animatable representations by construction. Together with a Motion-DiT model that operates on these structure-aware latents, RigMo enables controllable motion generation and seamless downstream applications. Experiments on multiple datasets show that RigMo discovers smooth, interpretable, and physically consistent rigs while outperforming existing auto-rigging and deformation methods in accuracy and generalization.

\newpage
{
    \small
    \bibliographystyle{ieeenat_fullname}
    \bibliography{main}
}

\clearpage
\setcounter{page}{1}
\maketitlesupplementary


\section*{Overview of Supplementary Materials}

\noindent
Due to the strict page limit of the main manuscript, several important details, analyses, and visualizations could not be presented. 
This supplementary document provides a comprehensive extension of our method and experiments. 
Specifically, we include the following additions:

\begin{itemize}[leftmargin=1.2em]
    \item \textbf{Comparison with prior work.}

    \item \textbf{RigMo-VAE.} 
    We provide side-by-side comparisons of the 48-token and 128-token configurations to highlight their differences in reconstruction quality.

    \item \textbf{MotionDiT.}
    We include a detailed description of the experimental settings, training configuration, dataset preparation, 
    and complete quantitative evaluations, accompanied by an in-depth analysis of the observed trends.

    \item \textbf{Video Demonstrations.}
    We further provide demo videos showcasing RigMo’s generated results (1-frame $\rightarrow$ 9-frame prediction), 
    along with side-by-side visual comparisons of reconstruction results against state-of-the-art baselines that could not fit in the main paper.
\end{itemize}

\vspace{4pt}
\noindent
Together, these supplementary materials offer a more complete view of RigMo’s effectiveness and support the claims made in the main manuscript.

\section*{Comparison with prior work.}
Existing approaches related to RigMo fall broadly into two categories: 
dynamic 3D generation methods and rigging–structure prediction methods. 
Despite impressive progress, none of these paradigms jointly model rigging and motion in a unified, 
self-supervised manner, which fundamentally limits their applicability to general deformable objects.

\textbf{Dynamic 3D generation.}
Recent 4D VAE models such as AnimateAnyMesh and GVFDiffusion learn trajectories for each vertex or 
Gaussian primitive. 
While effective for short-term sequence reconstruction, these models do not recover an explicit rig or 
kinematic structure, and typically require hundreds of latent tokens (often $512+$) to encode motion. 
In contrast, RigMo uses only $48$ latent tokens yet produces a compact, interpretable rig 
together with temporally coherent motion parameters.  
Another line of work extends static 3D VAEs, including Hunyuan3D and Step1X3D, to per-frame mesh 
generation. 
These methods operate on independent frames and therefore suffer from slow inference, lack of 
temporal smoothness, inconsistent geometry across time, and, most critically, no vertex 
correspondence between the generated shapes.  
Finally, motion–generation models designed for humans or articulated categories 
(e.g., AnyTop, many human motion diffusion/transformer models) assume access to ground-truth rigs.
Such assumptions break down for general objects where no consistent rig annotations exist, 
and rigging conventions vary widely across datasets and artists, resulting in poor cross-category 
generalization.

\textbf{Rigging prediction.}
Methods such as UniRig and Magic Articulate aim to predict rig structures from static geometry.  
These methods depend heavily on large-scale human-annotated rig datasets; however, manual rigging 
contains inconsistencies, annotation noise, and large variations across artists and modeling standards.  
Moreover, available datasets are small, category-specific, and difficult to scale, which limits 
generalization to diverse objects and motion patterns.  
RigMo, in contrast, requires no human annotation: it infers the underlying kinematic structure 
directly from observed motion.  
This motion-driven formulation yields rigging that is physically meaningful, consistent across 
instances, and automatically aligned with the actual deformation behavior of the object.

By unifying rigging and motion into a single generative model, RigMo overcomes the core limitations 
of both categories.  
It provides interpretable rigs, consistent mesh deformations, temporally stable trajectories, 
and strong generalization without relying on manual labels—offering a principled and scalable 
solution for general-object animation.

\begin{table}[t]
\centering
\renewcommand{\arraystretch}{1.12}
\setlength{\tabcolsep}{3.0pt}
\scalebox{0.7}{
\begin{tabular}{l|c|c|c}
\toprule
\multicolumn{4}{c}{\textbf{Dynamic 3D Generation}} \\
\midrule
Method & Explicit Rig & Temporal Consist. & Rig Annotation \\
\midrule
AnimateAnyMesh / GVFDiff. & \xmark & \checkmark & none \\
Hunyuan3D / Step1X3D & \xmark & \xmark & none \\
Human/AnyTop Motion Gen. & \checkmark (GT) & \checkmark & GT rigs \\
\midrule
\multicolumn{4}{c}{\textbf{Rigging Prediction}} \\
\midrule
UniRig / Magic Articul. & \checkmark & \xmark & GT rigs \\
\midrule
\multicolumn{4}{c}{\textbf{RigMo (Ours)}} \\
\midrule
RigMo & \checkmark & \checkmark & none \\
\bottomrule
\end{tabular}}
\caption{Comparison between RigMo and representative method families.}
\label{tab:rigmo_comparison}
\end{table}

\section*{MotionDiT Evaluation}

\begin{table*}[t]
\centering
\renewcommand{\arraystretch}{1.15}
\setlength{\tabcolsep}{3.8pt}  
\scalebox{0.8}{
\begin{tabular}{l|cc|cc}
\toprule
Method 
& \multicolumn{2}{c|}{1$\rightarrow$1} 
& \multicolumn{2}{c}{1$\rightarrow$9} \\
& Train & Val & Train & Val \\
& (L1/L2) & (L1/L2) & (L1/L2) & (L1/L2) \\
\midrule
w/o frame cond.
& 2.13$\pm$0.46 / 1.52$\pm$0.33 & 2.48$\pm$0.48 / 1.74$\pm$0.35
& 2.63$\pm$0.94 / 1.98$\pm$0.72 & 3.10$\pm$0.91 / 2.22$\pm$0.69 \\
latent rig cond.
& 1.89$\pm$0.42 / 1.31$\pm$0.32 & 2.01$\pm$0.43 / 1.45$\pm$0.31 
& 2.15$\pm$0.85 / 1.44$\pm$0.64 & 2.51$\pm$0.84 / 1.78$\pm$0.60 \\
full MotionDiT
& \textbf{1.43}$\pm$0.41 / \textbf{0.99}$\pm$0.30 
& \textbf{1.77}$\pm$0.41 / \textbf{1.23}$\pm$0.29
& \textbf{1.61}$\pm$0.83 / \textbf{1.18}$\pm$0.61
& \textbf{1.86}$\pm$0.82 / \textbf{1.37}$\pm$0.56 \\
\bottomrule
\end{tabular}}
\caption{
Ablation study of MotionDiT ($\times 10^{-2}$) under two sparse-conditioning settings.
We report Chamfer Distance (L1 / L2) for both training and validation sets.
The three variants include: 
(1) w/o frame condition (no frame-mask guidance), 
(2) latent rig condition (using only the rig-branch latent feature), and 
(3) full MotionDiT (latent rig feature + decoded skinning weights + Gaussian bone centers + frame-mask guidance). Experiments are done on DT4D test dataset.}
\label{tab:motiondit_eval}
\end{table*}

\begin{figure}[t]
    \centering
    \includegraphics[width=1\linewidth]{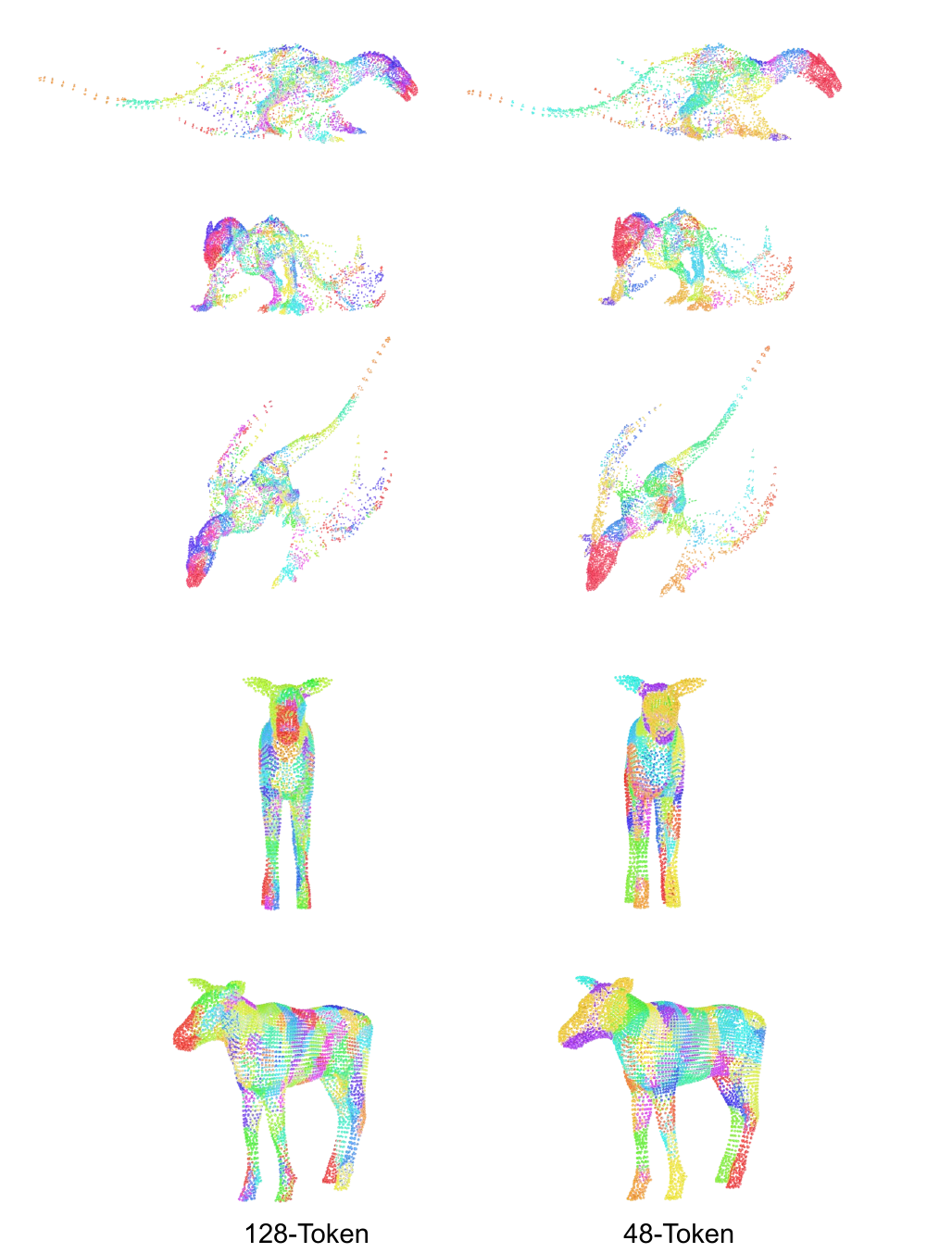}
    \vspace{0pt}
    \caption{Side-by-side skinning weights comparisons of the 48-token and 128-token configurations.}
    \vspace{0pt}
    \label{fig:6}
\end{figure}

\noindent
To more comprehensively evaluate the motion generation capability of MotionDiT, 
we design a set of controlled experiments that minimize sampling randomness and 
isolate the contribution of the diffusion-based temporal prediction module. 
Each experiment is conditioned on two complementary signals, described below.

\begin{itemize}[leftmargin=1.3em]
    \item Rigging condition:  
    MotionDiT receives rig-aware structural cues extracted from the RigMo-VAE rig branch.  
    These include the latent rig feature as well as the decoded, physically meaningful 
    rig attributes such as skinning weights and Gaussian bone centers, injected into 
    the first cross-attention layer (Fig.~3).  
    This conditioning provides the model with articulation structure and the expected 
    ranges of deformation.

    \item Frame condition (mask pattern):  
    A sparse set of observed frames is encoded into a frame-mask sequence and fed into 
    the second cross-attention layer.  
    This mask pattern controls which frames are visible to the model and enables 
    controlled interpolation and long-horizon prediction.
\end{itemize}

\noindent
We evaluate MotionDiT under two sparse-conditioning settings:

\begin{enumerate}[leftmargin=1.4em]
    \item 1-frame $\rightarrow$ 1-frame prediction:  
    Given a single input frame at time $t$, the model predicts the next frame at $t+1$, 
    focusing on local temporal smoothness and short-term motion fidelity.

    \item 1-frame $\rightarrow$ 9-frame prediction:  
    Given one observed frame, the model predicts the following nine frames.  
    This more challenging setting evaluates the ability of the model to produce 
    temporally coherent and physically plausible long-horizon motion.
\end{enumerate}

\noindent
These sparse-conditioning setups avoid free-form motion sampling and instead provide 
controlled evaluation of the temporal prediction capability of MotionDiT. 
They also allow a clean examination of the effect of different rig-conditioning strategies.

\medskip
\noindent
We compare three variants:

\begin{itemize}[leftmargin=1.3em]
    \item w/o frame condition:  
    Removes the frame-mask signal entirely, forcing the model to infer temporal structure 
    without any observed-frame guidance.

    \item Latent rig condition:  
    Conditions the model only on the latent rig feature extracted from the RigMo-VAE 
    rig encoder, without using decoded skinning weights or Gaussian bone centers.

    \item Full MotionDiT:  
    Uses the complete rig-conditioning bundle, including the latent rig feature, decoded 
    skinning weights, and Gaussian bone centers, providing explicit structural and 
    physically interpretable priors for motion prediction.
\end{itemize}

\noindent
Quantitative results on both the training and validation splits, measured using 
Chamfer Distance (CL$_1$ and CL$_2$), are summarized in Table~\ref{tab:motiondit_eval}.  
Across both prediction horizons, the full MotionDiT model achieves strong motion fidelity 
and temporal coherence.

\end{document}